\documentclass[final]{IEEEtran}
\usepackage{cite}
\usepackage{threeparttable}
\usepackage{graphicx}
\usepackage{picinpar}
\usepackage[cmex10]{amsmath}
\usepackage{amsmath,amsfonts,amssymb}
\usepackage{subfigure}

\usepackage{algorithm}
\usepackage{algorithmic}
\usepackage{ulem}

\begin{document}

\newtheorem{lemma}{Lemma}
\newtheorem{corol}{Corollary}
\newtheorem{theorem}{Theorem}
\newtheorem{proposition}{Proposition}
\newtheorem{definition}{Definition}
\newcommand{\e}{\begin{equation}}
\newcommand{\ee}{\end{equation}}
\newcommand{\eqn}{\begin{eqnarray}}
\newcommand{\eeqn}{\end{eqnarray}}

\title{Robust Preamble Design for Synchronization, Signaling Transmission and Channel Estimation}

\author{Zhen Gao~\IEEEmembership{Student Member,~IEEE}, Chao Zhang,~\IEEEmembership{Member,~IEEE},
 Zhaocheng Wang,~\IEEEmembership{Senior Member,~IEEE}%
\thanks{Manuscript received December 8, 2013; revised October 2, 2014; accepted
November 17, 2014. This work was supported by National Key Basic Research Program of China (Grant No. 2013CB329203),
National High Technology Research and Development Program of China (Grant No. 2014AA01A704),
and National Nature Science Foundation of China (Grant No. 61271266).}

\thanks{Z. Gao, C. Zhang, and Z. Wang are with Tsinghua National Laboratory for
 Information Science and Technology (TNList), Department of Electronic Engineering,
 Tsinghua University, Beijing 100084, China (E-mails: gao-z11@mails.tsinghua.edu.cn;
  z\_c@mail.tsinghua.edu.cn; zcwang@mail.tsinghua.edu.cn).} %
\vspace*{-5mm}
}

\maketitle

\begin{abstract}
 The European second generation digital video broadcasting standard (DVB-T2) introduces a P1 symbol. This P1 symbol facilitates the coarse synchronization and carries 7-bit transmission parameter signaling (TPS), including the fast Fourier transform size, single-input/single-output and multiple-input/single-output transmission modes, etc. However, this P1 symbol suffers from obvious performance loss over fading channels. In this paper, an improved preamble scheme is proposed, where a pair of optimal m sequences are inserted into the frequency domain. One sequence is used for carrier frequency offset (CFO) estimation, and the other carries TPS to inform the receiver about the transmission configuration parameters. Compared with the conventional preamble scheme, the proposed preamble improves CFO estimation performance and the signaling capacity. Meanwhile, without additional overhead, the proposed scheme exploits more active pilots than the conventional schemes. In this way, it can facilitate the channel estimation, improve the frame synchronization accuracy as well as enhance its robustness to frequency selective fading channels.
\end{abstract}


\IEEEpeerreviewmaketitle

\section{Introduction}

Orthogonal frequency division multiplexing (OFDM) has been widely adopted in broadband communication systems \cite{OFDM}. As a promising technology, OFDM has been applied in the European second generation digital video broadcasting standard (DVB-T2) and other digital television broadcasting solutions \cite{{DVB},{GB},{Dll},{ZC_EL},{Guideline},{dai_tds},{gao_vtc}}.

In order to accommodate various application scenarios like fixed, hand-held and mobile reception, DVB-T2 introduces a P1 symbol to convey 7-bit transmission configuration information, including the fast Fourier transform (FFT) size, single-input/single-output and multiple-input/single-output transmission modes, etc. Meanwhile, this P1 symbol also facilitates coarse timing synchronization and carrier frequency offset (CFO) estimation. Therefore, the reliability of P1 symbol detection and signaling demodulation is critical for the receiver to reliably perform its subsequent processing steps, e.g. channel estimation, data demodulation, etc.

In the P1 symbol, guard interval with shifted frequency is intended to mitigate continuous wave (CW) interference. However, this structure will deteriorate the preamble detection performance over multipath channels \cite{corre1}, \cite{my}. Meanwhile, an autocorrelation (denoted as trapezoid correlation) is adopted for preamble detection \cite{Guideline}, which cannot achieve good timing synchronization performance \cite{coarse}. Hence, an improved autocorrelation implementation (denoted as triangle correlation) is proposed to increase the timing accuracy \cite{friendly}. Unfortunately, the triangle correlation based preamble detection also suffers from obvious performance loss over multipath channels. For instance, two-tap unitary gain channel with 256 samples delay spread can entirely destroy the preamble detection \cite{my}.

In terms of signaling demodulation, a preamble based on distance detection (PBDD) is introduced \cite{PBDD} which has a better performance than the P1 symbol in DVB-T2. In this preamble, a pair of identical pseudo-noise (PN) sequences with variable distance are inserted in the frequency domain to convey the signaling information. However, active subcarriers arranged in a fraction of continuous block may result in the signaling demodulation failure when the channel gain of this continuous block suffers from severe deterioration in adverse circumstances.

In order to solve those problems, this paper proposes an improved preamble scheme, where a pair of optimal m sequences are inserted into the frequency domain to convey signaling and facilitate CFO estimation. At the receiver, a low-complexity circular correlation is used for CFO estimation and signaling demodulation. Compared with the conventional preamble schemes \cite{DVB}, \cite{PBDD}, the proposed preamble improves the CFO estimation performance and the signaling capacity. Meanwhile, in the proposed preamble scheme, all the subcarriers are exploited. In this way, it can facilitate the channel estimation, improve the accuracy of frame synchronization, and enhance its robustness to frequency selective fading channels.

The rest of the paper is organized as follows. Section II presents the conventional P1 symbol in DVB-T2 and its corresponding detection method. Section III proposes an improved preamble scheme. In Section IV, theoretical analyses are provided. Simulation results are given in Section V. Finally, conclusions are drawn in Section VI.
\section{P1 Symbol of DVB-T2}

\begin{figure}
     \centering
     \includegraphics[width=8.8cm, keepaspectratio]
     {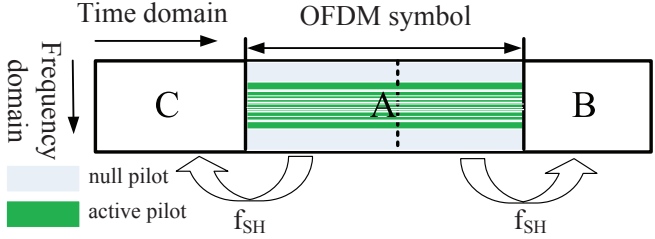}
     \caption{Structure of P1 symbol in DVB-T2.}
     \label{fig:STR_P1}
\end{figure}

In the time domain, as illustrated in Fig. \ref{fig:STR_P1}, the P1 symbol of DVB-T2 consists of three parts: C, A and B. Part ``A" is an OFDM symbol with FFT size $N=1024$, which carries 7-bit signaling. Part ``C" and part ``B", regarded as guard intervals, are the copies of the first 542 samples and the rest 482 samples of ``A" with a shifted frequency of ${f_{SH}}{\rm{ = }}1/(1024{T_s})$, respectively, where ${T_s}$ is the sample interval. Hence the P1 symbol in the time domain can be written in the following expression according to \cite{DVB}
\begin{align}
{p_n} = \left\{ \begin{array}{l}
{y_n} \cdot {e^{j2\pi {f_{SH}}n{T_s}}},{\kern 20pt}0 \le n < 542,\\
{y_{n - 542}},{\kern 49pt} 542 \le n < 1566,\\
{y_{n - 1024}} \cdot {e^{j2\pi {f_{SH}}n{T_s}}},1566 \le n < 2048,
\end{array} \right.
\end{align}
where $\{ {y_n}\} _{n = 0}^{N - 1}$ is the time domain baseband representation of part ``A".

In the frequency domain of ``A", only 384 subcarriers are active and others are set to be null. The distribution of active pilots is subject to a special pattern \cite{DVB}, and the active pilots are mainly arranged in the center of the frequency domain of the OFDM symbol, as shown in Fig. \ref{fig:STR_P1}. In addition, 7-bit signaling is divided into two signaling groups, i.e., 3-bit S1 and 4-bit S2. Patterns to encode S1 are based on 8 orthogonal sets from 8 complementary sequences of length 8 (total length of each S1 pattern is 64); while patterns to encode S2 are based on 16 orthogonal sets from 16 complementary sequences of length 16 (total length of each S2 pattern is 256). The S1 sequence, the S2 sequence, and a repetition of S1 sequence are concatenated to compose a length of 384 signaling sequence, which is first modulated by differential binary phase shift keying (DBPSK) and scrambled, and then mapped to active subcarriers.

For convenience, CFO can be normalized by the subcarrier spacing as the composition of the integer subcarrier spacing CFO part (IFO) and the fractional subcarrier spacing CFO part (FFO), i.e.,
\begin{align}
{f_{{\rm{CFO}}}} = {f_{{\rm{frac}}}} + {m_{{\rm{int}}}}\cdot{f_{{\rm{subcarrier}}}}
\end{align}
where FFO and subcarrier spacing are denoted as $f_{\text{frac}}$ and $f_{\text{subcarrier}}$, respectively, and $m_{\text{int}}$ is an integer. At the receiver, coarse synchronization and FFO estimation are initially obtained by the autocorrelation operation recommended by the implementation guidelines of DVB-T2 \cite{Guideline}. Afterwards, part ``A" is extracted from the received preamble and then transformed to the frequency domain with FFO compensation. To be specific, an energy-detection based subcarrier pattern matching is operated to recognize the active pilots, which may be circularly shifted due to the large CFO. In this way, IFO can be estimated from the received pattern location compared with the original location. Subsequently, the active carriers are descrambled and differential demodulated. Finally, the processed active subcarriers belonging to S1 and S2 are correlated with all the orthogonal complementary sequences, respectively. In this way, signaling S1 and S2 are detected.

However, the P1 symbol in DVB-T2 has several deficiencies. First, the autocorrelation based preamble detection cannot obtain good timing synchronization performance \cite{coarse}, \cite{friendly}. Second, the preamble structure of guard intervals with frequency shift will lead to the preamble detection performance deterioration over multipath fading channels \cite{corre1}, \cite{my}. In fact, a 0dB echo channel with certain delay spread will entirely destroy the preamble detection \cite{corre1}. Finally, the signaling demodulation exploits the correlation of adjacent active pilots. However, the correlation may be severely corrupted owing to some null pilots arranged between the adjacent active carriers \cite{PBDD}.
\section{Proposed Preamble}
In this section, without additional overhead, a novel preamble structure is proposed, and its corresponding detection algorithm will also be addressed.
\subsection{Structure of the Proposed Preamble}
\begin{figure}
     \centering
     \includegraphics[width=9cm, keepaspectratio]
     {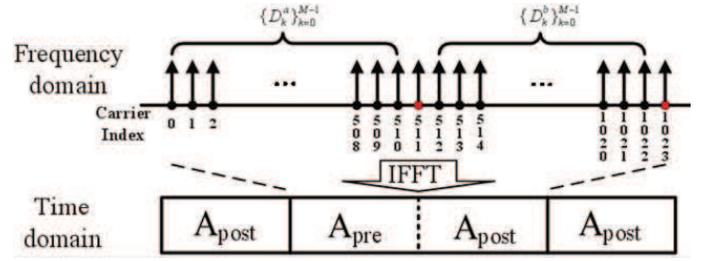}
     \caption{Structure of the proposed preamble.}
     \label{fig:mystr}
\end{figure}

The structure of the proposed preamble is shown in Fig. \ref{fig:mystr}. In the time domain, the proposed preamble is composed of an OFDM symbol with FFT size $N=1024$ and its prefix and postfix. The OFDM symbol is divided into two parts with equal length, which are denoted as $\text{A}_{\text{pre}}$ and $\text{A}_{\text{post}}$, respectively. $\text{A}_{\text{post}}$ is copied to the front and the end of the OFDM symbol as prefix and postfix, respectively. Hence, compared with the conventional preambles which use the structure of guard intervals with shifted frequency, the prefix in the proposed preamble can be regarded as the cyclic prefix (CP) of the OFDM symbol.

In the frequency domain of the OFDM symbol, all the pilots are active, which is different from the conventional schemes \cite{DVB}, \cite{PBDD}. The frequency domain of the OFDM symbol can be represented as
\begin{align}
{X_k} = \left\{ \begin{array}{l}
D_k^a,{\kern 1pt} {\kern 55pt}0 \le k < 511,\\
D_{\bmod (k + \Delta \phi  - 512,M)}^b,512 \le k < 1023,\\
 - 1,{\kern 60pt}k = 511{\kern 1pt} {\kern 1pt} {\rm{or}}{\kern 1pt} {\kern 1pt} 1023,
\end{array} \right.
\end{align}
where $\text{mod}(\cdot,M)$ is modulus $M$ operation, $\{ D_k^a\} _{k = 0}^{M - 1}$, $\{ D_k^b\} _{k = 0}^{M - 1}$ are a pair of length $M = 511$ sequence ($\pm $1 values), and $\Delta \phi  \in [0,510]$ is an integer, conveying the 9-bit signaling. Note that $\{ D_k^a\} _{k = 0}^{M - 1}$ and $\{ D_k^b\} _{k = 0}^{M - 1}$ meet the following properties, i.e.,
\begin{align}
D_k^c = D_k^a\cdot D_{\bmod (k + 1,M)}^a,0 \le k < M,\\
D_k^d = D_k^b\cdot D_{\bmod (k + 1,M)}^b,0 \le k < M,
\end{align}
where $\{ D_k^c\} _{k = 0}^{M - 1}$ and  $\{ D_k^d\} _{k = 0}^{M - 1}$ are a pair of optimum m sequence, which possesses the optimal cross correlation performance.

Specifically, subcarriers of index from 0 to 510 are determined by $\{ D_k^a\} _{k = 0}^{M - 1}$. These carriers enable the receiver to handle larger CFO. While carriers of index from 512 to 1022 are determined by $\{ D_k^b\} _{k = 0}^{M - 1}$ and signaling information, whereby signaling is distinguished by the circular shift phase of $\{ D_k^b\} _{k = 0}^{M - 1}$. In this way, the proposed preamble can convey 9-bit signaling information, which is larger than the conventional 7-bit signaling capacity. 	

\subsection{Detection Algorithm}

At the receiver, a adaptive detection method \cite{my} is adopted to improve the accuracy of coarse timing synchronization and eliminate CW interference. Afterwards, the OFDM symbol is extracted from the captured preamble and then transformed to the frequency domain by $N$-point FFT operation, i.e.,

\begin{align}
{\hat X_k} = \frac{1}{{\sqrt N }}\sum\limits_{n = 0}^{N - 1} {{{\hat r}_n}{e^{ - j2\pi {{\hat f}_{{\rm{frac}}}}n{T_s}}}{e^{ - j\frac{{2\pi }}{N}nk}}} ,0 \le k < N,
\end{align}
where ${\hat r_n}$ is the extracted time domain OFDM symbol, and ${\hat f_{{\rm{frac}}}}$ is the FFO estimation based on autocorrelation \cite{Guideline}. Subsequently, $\{ {\hat X_k}\} _{k = 0}^{N - 1}$ is binary differential demodulated, i.e.,
\begin{align}
{Y_k} = {\hat X_k}\cdot\hat X_{\bmod (k + 1,N)}^*,0 \le k < N,
\end{align}
where $(\bullet)^{*}$ is complex conjugation operation. After that, the differential demodulation output is circularly correlated with the local sequence $\{ D_k^c\} _{k = 0}^{M - 1}$, i.e.,
\begin{align}
R{m_l} = \frac{{\sum\limits_{k = 0}^{M - 1} {D{{_k^c}^*}{Y_{\bmod (k + l,N)}}} }}{{\sum\limits_{k = 0}^{M - 1} {|\hat X_{\bmod (k + l,N)}^{}{|^2}} }},0 \le l < N.
\end{align}
Hence the estimation of ${m_{{\rm{int}}}}$  can be written as
\begin{align}
{\hat m_{{\rm{int}}}} = \left\{ \begin{array}{l}
p,{\kern 1pt} {\kern 1pt} {\kern 1pt} {\kern 1pt} {\kern 1pt} {\kern 1pt} {\kern 1pt} {\kern 1pt} {\kern 1pt} {\kern 1pt} {\kern 1pt} {\kern 1pt} {\kern 1pt} {\kern 1pt} {\kern 1pt} {\kern 1pt} {\kern 1pt} {\kern 1pt} {\kern 1pt} {\kern 1pt} {\kern 1pt} {\kern 1pt} {\kern 1pt} {\kern 1pt} {\kern 1pt} {\kern 1pt} {\kern 1pt} {\kern 1pt} \quad 0 \le p < N/2,\\
p - N,\quad \;\;N/2 \le p < N,
\end{array} \right.
\end{align}
where $p$ is the index of correlation peak of (8), i.e.,
\begin{align}
p = \arg \mathop {\max }\limits_{0 \le l < N} |R{m_l}|.
\end{align}

After IFO correction, subcarriers of index from 512 to 1022 are extracted, denoted as $\{ {\hat C_k}\} _{k = 0}^{M - 1}$. Then $\{ {\hat C_k}\} _{k = 0}^{M - 1}$ is circularly correlated with $\{ D_k^d\} _{k = 0}^{M - 1}$, i.e.,
\begin{align}
R{\phi _q} = {\rm{ }}\frac{{\sum\limits_{k = 0}^{M - 1} {M{{_k^d}^*}\hat C_{\bmod (k + q,N)}^{}} }}{{\sum\limits_{k = 0}^{M - 1} {|\hat C_k^{}{|^2}} }},0 \le q < M.
\end{align}
Similarly, the estimation of $\Delta \phi$ can be expressed as
\begin{align}
\Delta \hat \phi  = \arg \mathop {\max }\limits_{0 \le q < M} |R{\phi _q}|.
\end{align}
Consequently, $\Delta \hat \phi $ is used to decode 9-bit signaling information.

\begin{figure}
     \centering
     \includegraphics[width=8.7cm, keepaspectratio]
     {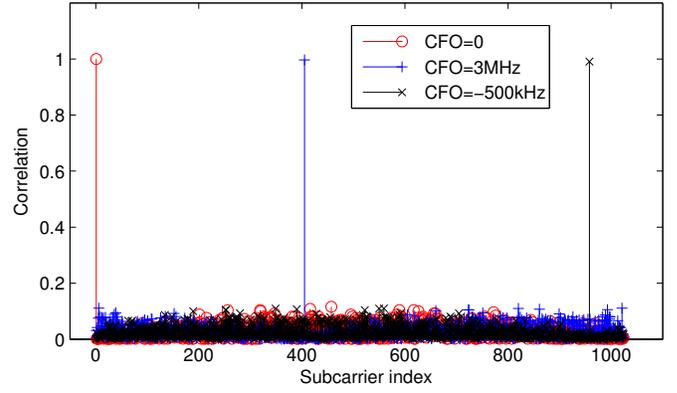}
     \caption{Correlation results of the IFO estimation in the proposed preamble over AWGN channel at SNR = 3dB.}
     \label{fig:IFO}
\end{figure}

\begin{figure}
     \centering
     \includegraphics[width=8.7cm, keepaspectratio]
     {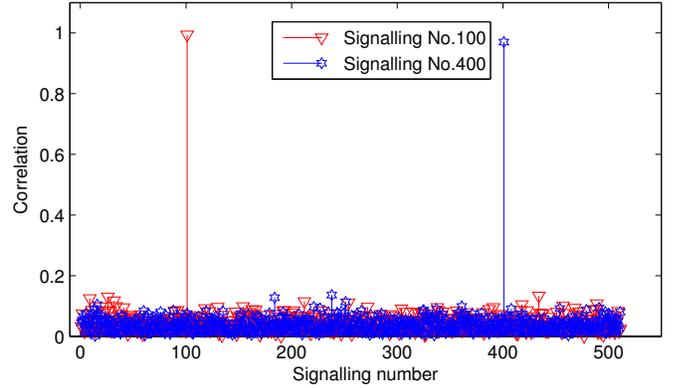}
     \caption{Correlation results of the signaling decoding in the proposed preamble over AWGN channel at SNR = 3dB.}
     \label{fig:corr}
\end{figure}

Examples of the correlation results of (8) and (11) recorded for transmission over the AWGN channel at SNR = 3dB are shown in Fig. \ref{fig:IFO} and Fig. \ref{fig:corr}, respectively. From Fig. \ref{fig:IFO} and Fig. \ref{fig:corr}, it can be observed that the proposed preamble can handle larger CFO and convey more signaling than the conventional preambles, and have good performance at low SNR.

Additionally, the circular correlation of (8) and (11) can be implemented by FFT with low complexity. For instance, in (11), 1024-point FFT/IFFT operation is implemented to calculate the 511-point circular correlation. In this FFT/IFFT operation, $\{ {\hat C_k}\} _{k = 0}^{M - 1}$ is repeated twice and padded two zeros to 1024 points, and $\{ D_k^d\} _{k = 0}^{M - 1}$ is padded 513 zeros to 1024 points. In this way, the complexity of length-511 circular correlation can be reduced.
\subsection{Channel Estimation and Accurate Synchronization}

Compared with the conventional preamble schemes which can facilitate coarse frame synchronization, estimate CFO and convey signaling, the proposed preamble can also be used for channel estimation as well as obtain accurate frame synchronization.

After CFO correction and signaling demodulation, the transmitted OFDM signaling can be regarded to be known at the receiver. Therefore, the channel frequency response can be estimated by any known criteria, such as least squares (LS) estimation \cite{LS}, i.e.,
\begin{align}
{\hat H_k} = {\hat X_k}/{X_k},0 \le k < N,
\end{align}
where $\left\{ {{{\hat H}_k}} \right\}_{k = 0}^{N - 1}$ is the frequency domain LS estimation of the wireless channels. It should be pointed out that, in this paper, we adopt the LS estimation due to its low complexity and simple implementation, and other advanced channel estimation methods such as minimum mean square error (MMSE) estimation can be also adopted to improve the performance of channel estimation.

The corresponding channel impulse response (CIR) can be expressed as \cite{time_frequency}
\begin{align}
{\hat h_n} = \frac{1}{{\sqrt N }}\sum\limits_{k = 0}^{N - 1} {{{\hat H}_k}{e^{j\frac{{2\pi }}{N}nk}}} ,0 \le n < N.
\end{align}
\begin{figure}
     \centering
     \includegraphics[width=8.5cm, keepaspectratio]
     {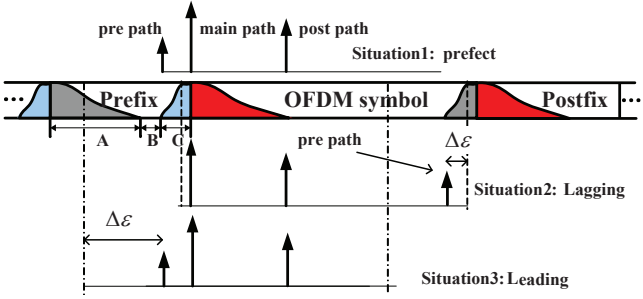}
     \caption{The impact of CIR estimation window.}
     \label{fig:fading}
\end{figure}

This CIR estimation plays a significant role in improving timing synchronization accuracy. Fig. \ref{fig:fading} depicts the impact of CIR estimation window position. Assuming perfect synchronization, the channel estimation window should start from the prefix without multipath contamination from previous data block, as the part B shown in Fig. \ref{fig:fading}. However, if the CIR estimation window lags, the pre-path may be regarded as the post-path in the estimated CIR; if the CIR estimation window leads, taps of CIR estimation will shift to the right. This is similar to the pre-path and post-path ambiguity problem in dual-PN OFDM (DPN-OFDM) system.

Thanks to the relatively longer OFDM symbol compared with the practical CIR length, when CIR estimation window position lags, the estimated CIR will be obvious longer than 512 samples, as the Situation2 shown in Fig. \ref{fig:fading}; when the CIR estimation window position leads, there is an obvious region without any active paths in the front of the estimated CIR, as the Situation3 shown in Fig. \ref{fig:fading}. Therefore, accurate timing synchronization can be achieved by exploiting the estimated CIR. The path delays of most significant taps are retained, i.e.,
\begin{align}
D = \{ n:|{\hat h_n}| \ge {p_{{\rm{th}}}}\} _{n = 0}^{N - 1},
\end{align}
where $p_{\text{th}}$ is the power threshold \cite{TH}. The first and last path index are denoted as
\begin{align}
{D_{{\rm{first}}}} = \min \{ D\} ,\\
{D_{{\rm{last}}}} = \max \{ D\} .
\end{align}
Thus, the synchronization sample offset based on perfect frame synchronization can be represented as
\begin{align}
\Delta \varepsilon  = \left\{ {\begin{array}{*{20}{c}}
{ - (N - {D_{{\rm{last}}}} + a), {D_{{\rm{last}}}} - {D_{{\rm{first}}}} > N/2,}\\
{{D_{{\rm{first}}}} - a,\;\;\;\;\;\;\;\;\;\;{D_{{\rm{last}}}} - {D_{{\rm{first}}}} \le N/2,}
\end{array}} \right.
\end{align}
where $a$ is a positive number used to combat interference in case that some low power active paths may be treated as noise. $\Delta \varepsilon  > 0$ means that CIR estimation window position need to be lagged, and vice versa.

The CIR estimation (13), (14) can also be used for the following OFDM symbols. For instance, a P2 symbol following the P1 symbol can provide static, configurable, dynamic layer-1 signaling as well as data in DVB-T2 system \cite{DVB}. To enhance its robustness, P2 symbol is repeated more than once and extra pilots are inserted \cite{DVB}. Therefore, the proposed preamble with coarse channel estimation can improve the P2 symbol demodulation or reduce the P2 symbol repetition number.
\section{Theoretical Performance Analyses}
This section focuses on the theoretical performance analyses of IFO estimation, signaling demodulation, etc., over additive white Gaussian noise (AWGN) channel.
\subsection{Distribution and Detection Probability of IFO}
Assuming perfect timing synchronization and FFO correction, under AWGN channel, (6) can be written as ${\hat X_k} = {X_k} + {W_k}$, where ${W_k}$ is AWGN in the frequency domain. Thus the numerator and denominator in (8) can be written as \cite{CFO_detect}
\begin{align}
&N{m_l} = \sum\limits_{k = 0}^{M - 1} {D{{_k^c}^*}{Y_{\bmod (k + l,N)}}} ,0 \le l < N,\\
&D{m_l} = \sum\limits_{k = 0}^{M - 1} {|{{\hat X}_{\bmod (k + l,N)}}{|^2}} ,0 \le l < N.
\end{align}

Thus (8) can be written as $Rm_l=Nm_l/Dm_l$. When $l=m_{\text{int}}$, namely the correlation peak, the numerator can be written as
\begin{align}
N{m_{{\rm{peak}}}} = \sum\limits_{k = 0}^{M - 1} {\underbrace {\sigma _s^2|D_k^c{|^2}}_{{\rm{constant}}{\kern 1pt} {\kern 1pt} {\rm{ = }}{\kern 1pt} {\kern 1pt} M\sigma _s^2} + {\sigma _s}D_{k + 1}^aW_{k + 1}^*} \nonumber \\
+{\sigma _s}D_k^{a*}{W_k}+ D_{k + 1}^aD_k^{a*}{W_k}W_{k + 1}^*
\end{align}
where, $\sigma _s^2$ and $\sigma _w^2$ denote the power of signal and AWGN, respectively. According to the central limit theorem, $Nm_{\text{peak}}$ is approximately subjected to complex Gaussian distribution, and the real part and the imaginary part of $Nm_{\text{peak}}$ are also subjected to Gaussian distributions and independent of each other \cite{effi}, i.e.,
\begin{align}
&{\rm{Re}}\{ N{m_{{\rm{peak}}}}\}  \sim  N(M\sigma _s^2,(2M\sigma _s^2\sigma _w^2 + M\sigma _w^4)/2),\\
&{\mathop{\rm Im}\nolimits} \{ N{m_{{\rm{peak}}}}\} \sim N(0,(2M\sigma _s^2\sigma _w^2 + M\sigma _w^4)/2).
\end{align}

Outside the correlation peak, the numerator of (8), denoted as $Nm_{\text{out}}$, can be written as
\begin{align}
\!\!\!\!\!\!&N{m_{{\rm{out}}}} = \underbrace {\sum\limits_{k = 0}^{M - 1} {\sigma _S^2D_k^c\frac{{{X_k}}}{{|{X_k}|}}\frac{{X_{k + 1}^*}}{{|X_{k + 1}^*|}}} }_{{\rm{Item1}} \approx {\rm{0}}}\nonumber  \\
& + \underbrace {\sum\limits_{k = 0}^{M - 1} {{\sigma _s}D_k^c\frac{{{X_k}}}{{|{X_k}|}}W_{k + 1}^* + {\sigma _s}D_k^c\frac{{X_{k + 1}^*}}{{|X_{k + 1}^*|}}{W_k}{\rm{ + }}D_k^c{W_k}W_{k + 1}^*} }_{{\rm{Item2}}}.
\end{align}
In (24), the sequence $\{ D_k^c{X_k}{X_{k + 1}}/|{X_k}||{X_{k + 1}}|\} _{k = 0}^{M - 1}$ is a determined pseudo-random sequence and its sum approximates to zero. So Item1 can be ignored and $Nm_{\text{out}}$ follows the distribution
\begin{align}
{\rm{Re}}\{ N{m_{{\rm{out}}}}\} \sim N(0,\frac{{M(\sigma _w^4 + 2\sigma _s^2\sigma _w^2)}}{2})\\
{\mathop{\rm Im}\nolimits} \{ N{m_{{\rm{out}}}}\} \sim N(0,\frac{{M(\sigma _w^4 + 2\sigma _s^2\sigma _w^2)}}{2})
\end{align}
Similarly, $Dm_{l}$ follows
\begin{align}
&{\rm{Re}}\{ D{m_l}\} \sim N(M(\sigma _w^2 + \sigma _s^2),M(\sigma _w^4 + 2\sigma _w^2\sigma _s^2))\\
&{\rm{Im}}\{ D{m_l}\} \sim N(0,2M\sigma _w^2\sigma _s^2))
\end{align}
The real part and the imaginary part of $Dm_l$ and $Nm_{\text{out}}$ are also independent of each other, respectively. Thus ${\left| {N{m_{{\rm{peak}}}}} \right|^2}$, ${\left| {D{m_l}} \right|^2}$ and ${\left| {N{m_{{\rm{out}}}}} \right|^2}$ follow chi-square distribution with freedom degree of 2, where the first two are non-central \cite{tool}. Hence ${\left| {R{m_{{\rm{peak}}}}} \right|^2} = {\left| {N{m_{{\rm{peak}}}}} \right|^2}/{\left| {R{m_l}} \right|^2}$ and ${\left| {R{m_{{\rm{out}}}}} \right|^2} = {\left| {N{m_{{\rm{out}}}}} \right|^2}/{\left| {R{m_l}} \right|^2}$ follow $F_{v1,v2}$ distribution, where $v1=v2=2$ \cite{tool}. Since the mean of $Dm_{l}$ is much larger than its variance in the practical signal-to-noise ratio (SNR) range, it can be approximated as
\begin{align}
&|R{m_{{\rm{peak}}}}{|^2} \approx \frac{{\sigma _{{\mathop{\rm Re}\nolimits} \{ N{m_{{\rm{peak}}}}\} }^2}}{{\mu _{{\mathop{\rm Re}\nolimits} \{ D{m_l}\} }^2}}\chi _v^2({\lambda _1})\\
&|R{m_{{\rm{out}}}}{|^2} \approx \frac{{\sigma _{{\mathop{\rm Re}\nolimits} \{ N{m_{{\rm{out}}}}\} }^2}}{{\mu _{{\mathop{\rm Re}\nolimits} \{ D{m_l}\} }^2}}\chi _v^2({\lambda _2})
\end{align}
where, $v=2$, ${\lambda _1} = \mu _{{\rm{Re}}\{ D{m_{{\rm{peak}}}}\} }^2{\rm{ }}/\sigma _{{\rm{Re}}\{ D{m_{{\rm{peak}}}}\} }^2{\rm{ }}$, ${\lambda _2} = 0$. ${\mu _ \bullet }$ and $\sigma _ \bullet ^2$ mean the expectation and variance of a random variable, respectively.

From (29) and (30), the probability density function (PDF) of ${\left| {R{m_{{\rm{peak}}}}} \right|^2}$ and ${\left| {R{m_{{\rm{out}}}}} \right|^2}$ are denoted as $f_{\text{peak}}$ and $f_{\text{out}}$, respectively, i.e.,
\begin{align}
{f_{{\rm{peak}}}}(x) = \left\{ {\begin{array}{*{20}{c}}
\begin{array}{l}
\frac{1}{{2{u_1}}}{e^{ - \frac{1}{2}(x/{u_1} + {\lambda _1})}}{I_0}(\sqrt {{\lambda _1}x/{u_1}} ),\;x > 0\\
0,\;\;\;\;\;\;\;\;\;\;\;\;\;\;\;\;\;\;\;\;\;\;\;\;\;\;\;\;\;\;\;\;\;\;\;\;\;x < 0
\end{array}
\end{array}} \right.
\end{align}
\begin{align}
{f_{{\rm{out}}}}(y) = \left\{ {\begin{array}{*{20}{c}}
{\frac{1}{{2{u_2}}}{e^{ - \frac{1}{{2{u_2}}}y}},y > 0}\\
{0,\;\;\;\;\;\;\;\;\;\;\;y < 0}
\end{array}} \right.
\end{align}
where, ${u_1} = \sigma _{{\rm{Re}}\{ N{m_{{\rm{peak}}}}\} }^2/\mu _{{\rm{Re}}\{ D{m_l}\} }^2$, ${u_2} = \sigma _{{\rm{Re}}\{ N{m_{{\rm{out}}}}\} }^2/\mu _{{\rm{Re}}\{ D{m_l}\} }^2$, and $I_0(\bullet)$ is zero-order modified Bessel function of the first kind. Furthermore, given $z = {\rm{max|}}R{m_{{\rm{out}}}}{|^2} = \mathop {\max }\limits_{0 \le l < N,l \ne {m_{{\rm{int}}}}} |R{m_l}{|^2}$, its PDF is
\begin{align}
\begin{array}{l}
{f_{{\rm{ma}}{{\rm{x}}_{{\rm{out}}}}}}(z) = (N - 1){(\int_{ - \infty }^z {{f_{{\rm{out}}}}(z)dz} )^{N - 2}}{f_{{\rm{out}}}}(z)\\
{\kern 1pt} {\kern 1pt} {\kern 1pt} {\kern 1pt} {\kern 1pt} {\kern 1pt} {\kern 1pt} {\kern 1pt} {\kern 1pt} {\kern 1pt} {\kern 1pt} {\kern 1pt} {\kern 1pt} {\kern 1pt} {\kern 1pt} {\kern 1pt} {\kern 1pt} {\kern 1pt} {\kern 1pt} {\kern 1pt} {\kern 1pt} {\kern 1pt} {\kern 1pt} {\kern 1pt} {\kern 1pt} {\rm{     }} = \left\{ {\begin{array}{*{20}{c}}
{\frac{{N - 1}}{{2{u_2}}}{{(1 - {e^{ - \frac{1}{{2{u_2}}}z}})}^{N - 2}}{e^{ - \frac{1}{{2{u_2}}}z}}{\kern 1pt} {\kern 1pt} {\kern 1pt} {\kern 1pt} {\kern 1pt} {\kern 1pt} {\kern 1pt} {\kern 1pt} {\kern 1pt} {\kern 1pt} {\kern 1pt} {\kern 1pt} {\kern 1pt} {\kern 1pt} {\kern 1pt} z \le 0}\\
{0\;\;\;\;\;\;\;\;\;\;\;\;\;\;\;\;\;\;\;\;\;\;\;\;\;\;\;\;\;\;\;\;\;\;\;\;\;\;z \le 0}
\end{array}} \right.
\end{array}
\end{align}
Thus,
\begin{align}
\!\!\!\!\!\!\!\!\!\!\!\!\!\!\!\!&P(\{ \mathop {\max }\limits_{0 \le l < N,l \ne {m_{{\rm{int}}}}} |R{m_l}{|^2}\}  > |R{m_{{m_{{\rm{int}}}}}}{|^2}) = \frac{1}{{2{u_1}}}\exp ( - \frac{{{\lambda _1}}}{2})\nonumber \\
& \times \int\limits_0^{ + \infty } {[1 - {{(1 - \exp ( - \frac{1}{{2{u_1}}}z))}^{N - 1}}]\exp ( - \frac{x}{{2{u_1}}})} {I_0}(\sqrt {\frac{{{\lambda _1}x}}{{{u_1}}}} )dx
\end{align}
Namely, the false IFO probability can be written as
\begin{align}
{P_{{\rm{false}}\_{\rm{IFO}}}} = P(\{ \mathop {\max }\limits_{0 \le l < N,l \ne {m_{{\rm{int}}}}} |R{m_l}{|^2}\}  > |R{m_{{m_{{\rm{int}}}}}}{|^2}).
\end{align}
\subsection{Detection Probability of Signaling Demodulation}
Assuming the perfect IFO, the distribution analysis of $R\phi$ is similar to that of $Rm$. According to the circular correlation performance of m sequence, when sequence phase mismatches and the sequence length is large enough, the circular correlation value is approximated to 0. Hence, the conclusions of $Rm$ (29), (30) are also appropriate to $R\phi$. Thus, the error signaling demodulation probability is
\begin{align}
{P_{{\rm{err}}\_{\rm{sig}}}} = P(\{ \mathop {\max }\limits_{0 \le q < N,q \ne \Delta \phi } |R{\phi _q}{|^2}\}  > |R{\phi _{\Delta \phi }}{|^2}).
\end{align}
Consequently, considering the IFO, the final false signaling demodulation probability is
\begin{align}
{P_{{\rm{flase}}\_{\rm{sig}}}} = 1 - (1 - {P_{{\rm{false}}\_{\rm{IFO}}}})(1 - {P_{{\rm{err}}\_{\rm{sig}}}}).
\end{align}
\subsection{Cramer Rao Low Bound of Channel Estimation}
The CRLB of the channel estimation \cite{Dai} can be expressed as
\begin{align}
{\rm{CRLB = E}}\{ {\left. {\left\| {\hat h - h} \right.} \right\|_2}\} {\rm{ = }}\frac{{{L_h}}}{{N\rho }},
\end{align}
where, $\hat h$ is the estimated channel, $L_h$ is the channel length. Assuming the channel length is confined to 512 samples delay spread, thus $L_h=512$, and $N=1024$.
\section{Simulation Results}
\begin{table}[tb]%
\setbox0\hbox{\verb/\documentclass/}%
\caption{Channel Parameters}
\label{classoption}
\begin{center}
\begin{tabular}{c |c | c|c }
\hline
\hline
  &\multicolumn{1}{c|}{ITU-VB} & \multicolumn{1}{c|}{CDT-8} & \multicolumn{1}{c}{BSC} \\
 \hline
Echo &Delay/Gain/Phase & Delay/Gain/Phase & Delay/Gain/Phase  \\
Tap &($\mu{s}$)/(dB)/(rad) & ($\mu{s}$)/(dB)/(rad) & ($\mu{s}$)/(dB)/(rad)\\
\hline
0 &0.00 / -2.5/0 & -1.80/-18.0/0 & 0.1314/ -18.8500/0 \\
1 &0.30 /  0.0/0 & 0.00 /  0.0/0 & 0.6570/ -13.8471/$\pi$   \\
2 &8.90 /-12.8/0 & 0.15 /-20.0/0 & 1.1827/  -4.0248/0 \\
3 &12.90/-10.0/0 & 1.80 /-20.0/0 & 1.4455/   0.0000/0  \\
4 &17.10/-25.2/0 & 5.70 /-10.0/0 & 1.7083/  -4.0248/0 \\
5 &20.00/-16.0/0 & 30.00/  0.0/0 & 2.2339/ -13.8471/$\pi$   \\
6 &              &               & 2.7595/ -18.8500/0     \\
 \hline  \hline
\end{tabular}
\end{center}
\end{table}

In this section, signaling error rate (SER), IFO error rate (IFOER), timing error and channel estimation performance are investigated. Simulations adopt the same transmission parameters with DVB-T2 scenario, and the signal powers of all the preambles are normalized in the simulations to ensure a fair comparison. Meanwhile, AWGN channel, International Telecommunication Union Vehicular-B (ITU-VB) channel \cite{ITU_VB}, the China digital television test 8th channel model (CDT-8) \cite{CDT_8}, and a band-stop channel (BSC) are adopted. The detailed channel parameters are shown in TABLE I. In this section, for Fig. 6 and Fig. 9, we compute the SER/IFOER after the number of signalling errors or IFO errors is larger than 100 (e.g., if the SER is $10^{-5}$, the number of trials is $10^{7}$), while the number of trials for each Monte Carlo simulation in Fig. 10 and Fig. 11 is $10^4$.

\begin{figure}[tb]
     \begin{center}
     \includegraphics[width=8.7cm, keepaspectratio]{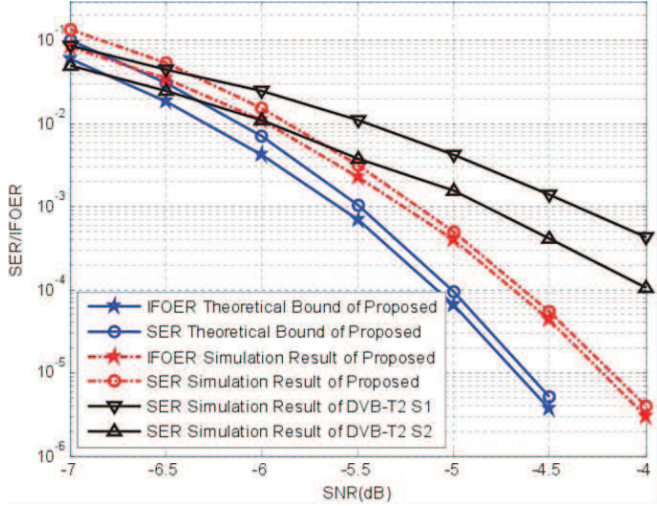}
     \end{center}
     \vspace*{-2mm}
     \caption{SER and IFOER simulation results of the proposed preamble compared with theoretical bound over AWGN channel.}
     \label{fig:awgn_ser}
     \vspace*{-4mm}
\end{figure}

Fig. \ref{fig:awgn_ser} depicts SER and IFOER simulation results of the proposed preamble compared with the theoretical analyses (35), (37) over AWGN channel. From Fig. \ref{fig:awgn_ser}, it is clear that the performance curves obtained via the theoretical approach are in good agreement with that of the Monte Carlo simulation results, and the gap between the Monte Carlo simulation results and the theoretical bound is around 0.5 dB when the target SER/IFOER of $10^{-5}$ is considered. Such gap between the Monte Carlo simulation results and the theoretical bound is because the theoretical bound of SER/IFOER provided in Section IV is based on the ideal assumption of the perfect timing synchronization and accurate FFO correction, while in the practical preamble detection, the coarse timing synchronization is based on the autocorrelation operation which is not accurate, and the FFO has to be estimated and compensated, thus the achievable SER/IFOER will suffer from a slightly performance loss compared with the theoretical bound. Additionally, the SERs of S1 and S2 of the P1 symbol are also plotted for comparison, which are inferior to the proposed preamble.
\begin{figure}[t]
     \begin{center}
     \includegraphics[width=8.9cm, keepaspectratio]{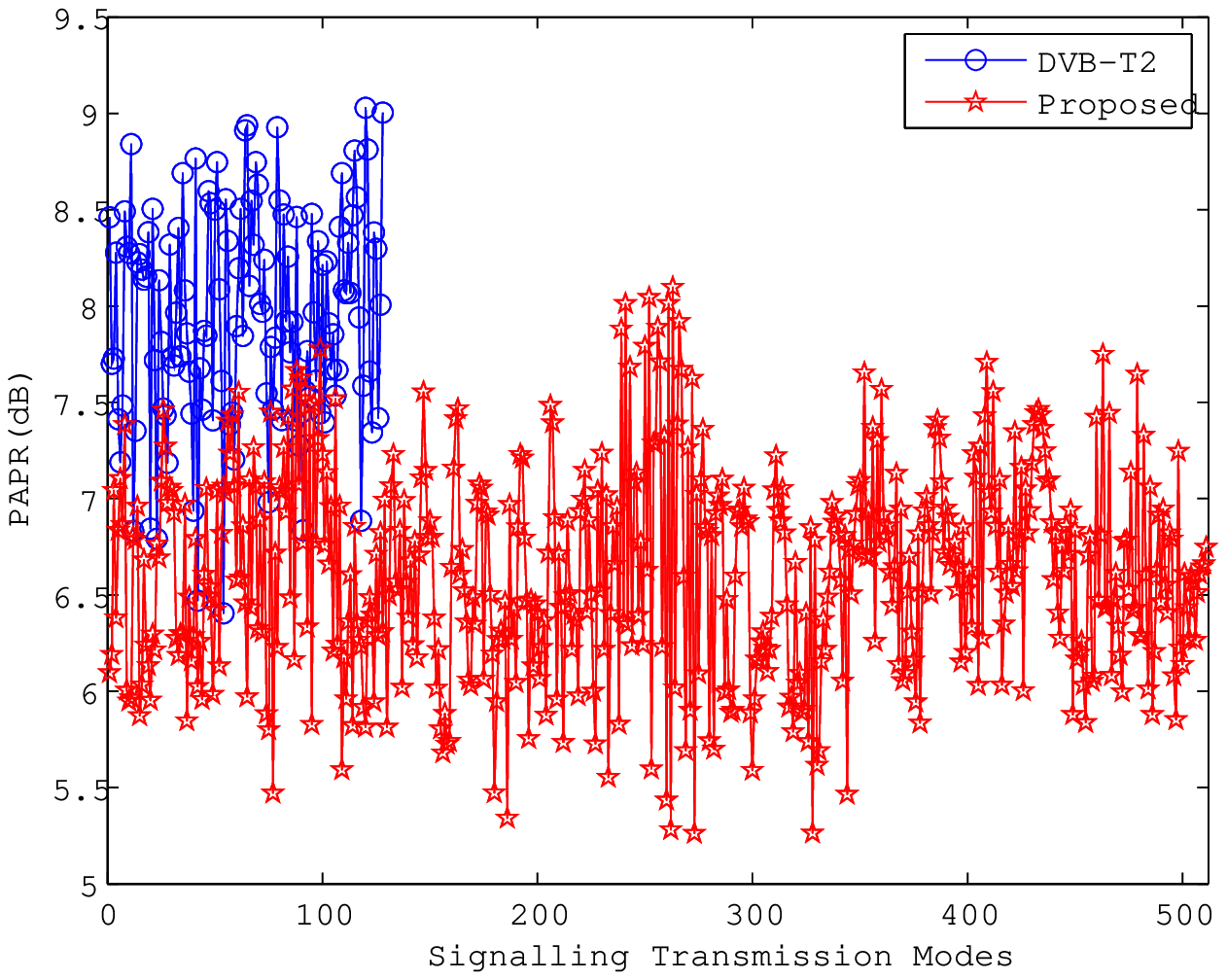}
     \end{center}
     \caption{PAPR comparison of the proposed preamble with 511 signal transmission modes and the conventional P1 symbol in DVB-T2 with 128 signal transmission modes.}
     \label{fig:papr1}
     \vspace*{1mm}
\end{figure}

\begin{figure}[t]
     \begin{center}
     \includegraphics[width=8.5cm, keepaspectratio]{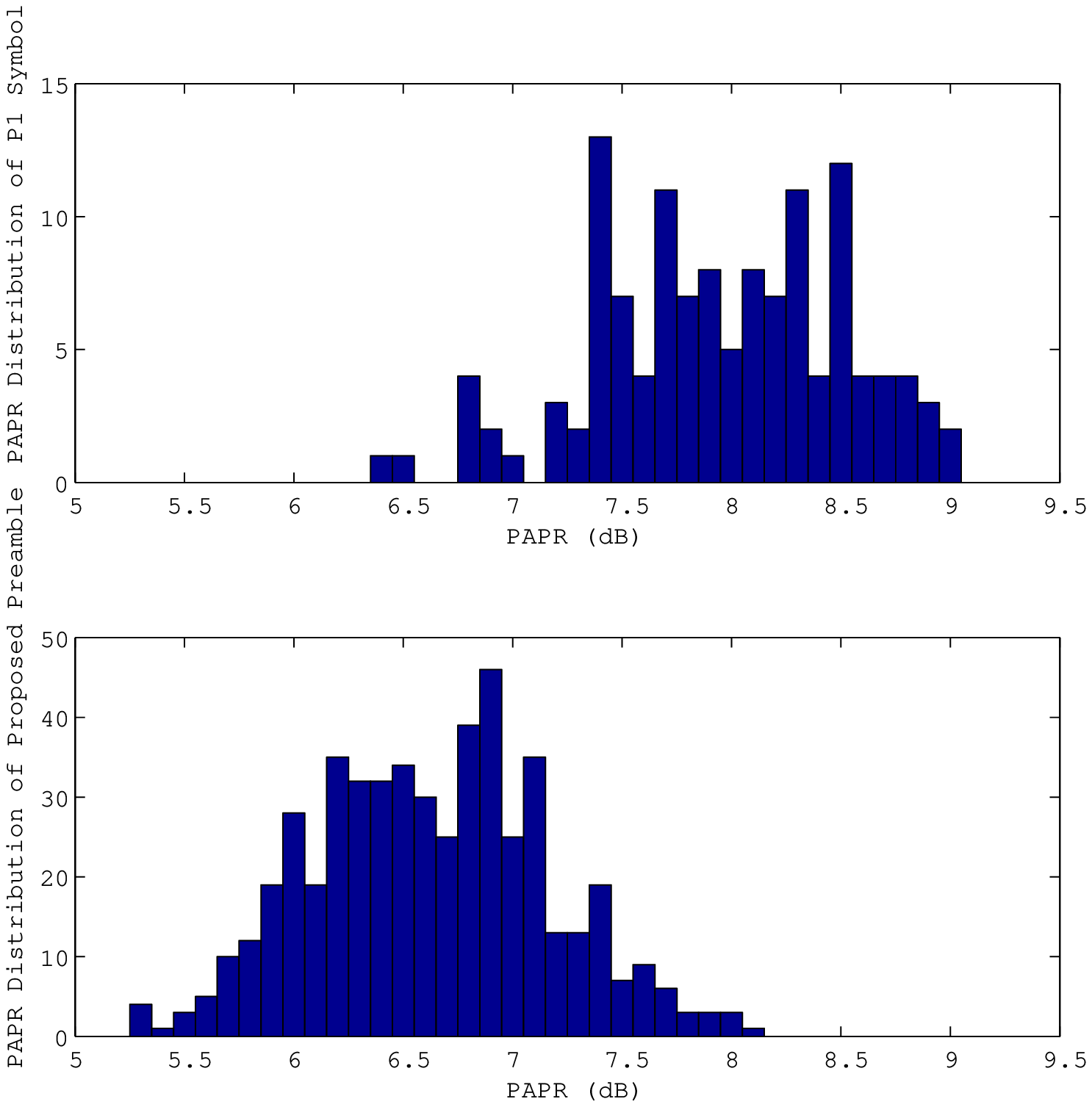}
     \end{center}
     \caption{Comparison of the PAPR distribution of the proposed preamble with 511 signal transmission modes and the conventional P1 symbol in DVB-T2 with 128 signal transmission modes.}
     \label{fig:papr2}
\end{figure}

Fig. \ref{fig:papr1} and Fig. \ref{fig:papr2} compare the peak-to-average power ratio (PAPR) of the proposed preamble with the P1 symbol in DVB-T2. Here PAPR of a deterministic signal can be defined as~\cite{PAPR}
\begin{align}
{\rm{PAPR = 10lo}}{{\rm{g}}_{10}}(\frac{{\mathop {\max }\limits_n |{s_n}{|^2}}}{{E\{ |{s_n}{|^2}\} }}),
\end{align}
where $E{\rm{\{\bullet }}{\rm{\} }}$ is the expectation operation, and $\left\{ {{s_n}} \right\}_{n = 0}^{{N_s} - 1}$ is a deterministic signal with the length of $N_s$.
Fig. \ref{fig:papr1} plots the PAPR of 511 signal transmission modes of the proposed preamble scheme and 128 signal transmission modes of the conventional P1 symbol in DVB-T2, where the horizontal axis of Fig. \ref{fig:papr1} denotes the signalling transmission modes and the vertical axis denotes the PAPR of the signalling with the corresponding transmission mode. From Fig. \ref{fig:papr1}, it is clear that most modes of the proposed preamble scheme enjoy the lower PAPR than that of the P1 symbol in DVB-T2.
In order to compare the PAPR of two preambles more intuitively, Fig. \ref{fig:papr2} compares the PAPR distribution of two preambles.
From Fig. \ref{fig:papr2}, it can be found that the PAPR of most signalling transmission modes of P1 symbol is between 7 dB and 9 dB. In contrast, the PAPR of most signalling transmission modes of the proposed preamble scheme is between 5.5 dB and 7.5 dB. Consequently, for most of the signalling modes, the proposed preamble has the lower PAPR than that of P1 symbol.

\begin{figure}[tb]
     \begin{center}
     \includegraphics[width=8.7cm, keepaspectratio]{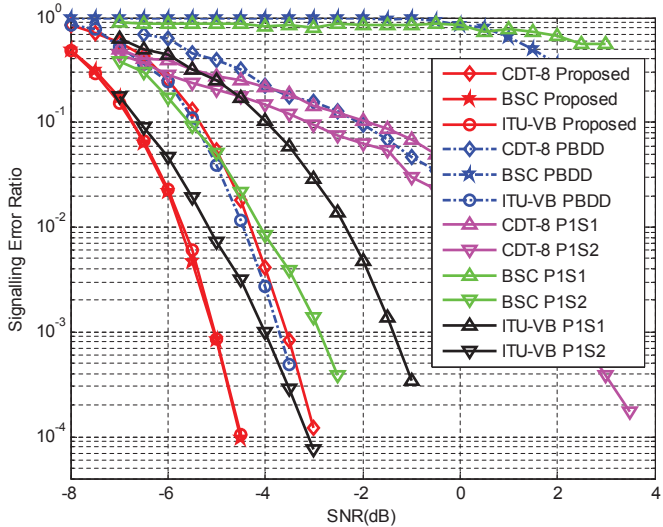}
     \end{center}
     \caption{SER performance comparison over multipath channels, where P1 symbol is divided into S1 and S2 denoted as ``P1S1", ``P1S2", respectively.}
     \label{fig:ser_fading}
\end{figure}
Fig. \ref{fig:ser_fading} shows the SER performance of the proposed scheme and its conventional counterparts: the P1 symbol and PBDD\footnote{In this paper, PBDD adopts $f_{SH}=1/(512T_s)$ in its time-domain structure according to [8], which is different from $f_{SH}=1/(1024T_s)$ in PBDD originally proposed in [11]. The reason is that $f_{SH}=1/(512T_s)$ can achieve a better elimination of the CW interference since PBDD adopts the coarse synchronization proposed in [10], while $f_{SH}=1/(1024T_s)$ cannot effectively combat the CW interference when the coarse synchronization proposed in [10] is adopted.} under selected multipath channels. Note that, in DVB-T2 scenario, CDT-8 is a channel of 243 samples delay spread with a 0dB echo, ITU-VB is a channel of 153 samples delay spread, and BSC is a channel of 21 samples delay spread with obvious band-stop characteristic. From Fig. \ref{fig:ser_fading}, it can be observed that, compared with the proposed preamble, PBDD fails to work in BSC, and suffers from an obvious performance loss over CDT-8 and ITU-VB channel. The performance of S1 and S2 of the P1 symbol is not satisfactory. However, the proposed scheme performs much better than the conventional preamble schemes.
\begin{figure}[tb]
     \begin{center}
     \includegraphics[width=8.9cm, keepaspectratio]{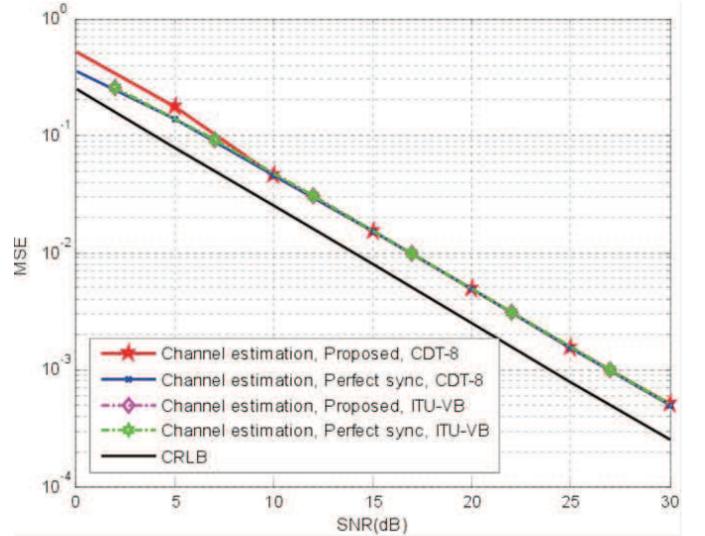}
     \end{center}
      \vspace*{-2mm}
     \caption{Channel estimation of the proposed preamble.}
     \label{fig:ce}
\end{figure}
\begin{figure}[tb]
     \begin{center}
     \includegraphics[width=8.9cm, keepaspectratio]{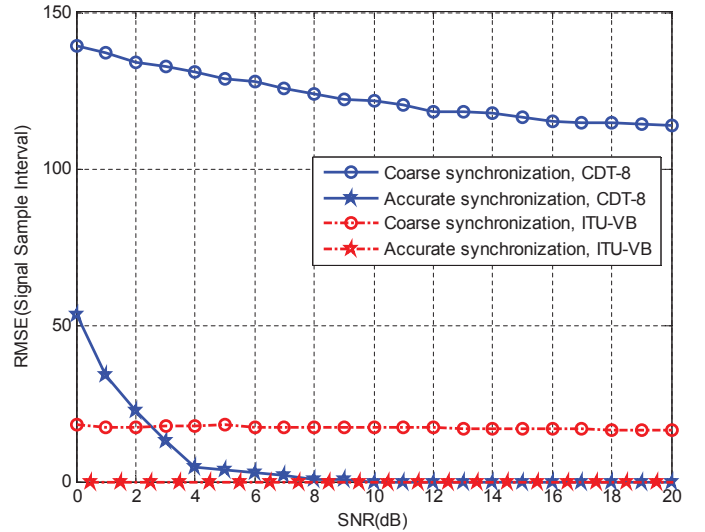}
     \end{center}
      \vspace*{-2mm}
     \caption{Timing synchronization performance of the proposed preamble.}
     \label{fig:sync}
\end{figure}

The reason for the much better signaling performance of the proposed preamble for transmission over severely frequency-selective channels can be explained as follows. First, the proposed scheme exploits all the pilots in frequency domain, which obtains frequency diversity. In this way, the proposed preamble is more robust to fading channels, compared with the conventional preambles which only exploit part of the pilots. Additionally, the guard interval meets the cyclic prefix characteristic for the OFDM symbol, which increases the tolerance of imperfect frame synchronization. Hence for channels like CDT-8 which may result in relative large frame synchronization offset, the proposed scheme still works well.

Fig. \ref{fig:ce} depicts the channel estimation mean square error (MSE) of the proposed preamble over CDT-8 and ITU-VB channels. The channel estimation MSE over BSC channel is not plotted for comparison, since it has a close performance to that of ITU-VB. The channel estimation MSE based on perfect frame synchronization and CRLB are also included for comparison. From Fig. \ref{fig:ce}, it can be observed that the channel estimation of the proposed preamble performs closely to the channel estimation based on perfect frame synchronization, and the performance loss from the theoretical bound CRLB mainly results from the imperfect FFO.

Fig. \ref{fig:sync} depicts timing synchronization performance of the proposed preamble over CDT-8 and ITU-VB channels. The timing synchronization performance over BSC channel is not plotted for comparison, since it has a similar performance to that of ITU-VB. Frame synchronization offset is measured by root-mean-square error (RMSE). RMSE is defined as $\sqrt {\frac{1}{S}\sum\limits_{s = 1}^S {{{\left( {{{\hat P}_s} - {P_{{\rm{ideal}}}}} \right)}^2}} } $, where $S$ is the number of simulation times, and ${P_{{\rm{ideal}}}}$ and ${{{\hat P}_s}}$ are the ideal and estimated frame synchronization position, respectively. Fig. \ref{fig:sync} indicates that the accurate synchronization of the proposed preamble improves the frame synchronization performance tremendously, especially over multipath channels with long delay spread e.g. CDT-8. Meanwhile, for other channels e.g. ITU-VB, the accurate synchronization of the proposed preamble scheme can even achieve perfect frame synchronization at SNR=0dB. It should be pointed out that if we adopt some other advanced channel estimation schemes such as the MMSE based channel estimator, the performance of accurate timing synchronization can be improved further.
\section{Conclusion}
Without extra overhead, a novel preamble was proposed in this paper. Compared with the conventional preamble schemes, the proposed solution improves the signaling capacity by 2 bit and enlarges the IFO estimation range further. Meanwhile, in contrast with other state-of-art preambles, the proposed preamble exploits all the pilots in the frequency domain. In this way, the proposed preamble is more robust than the conventional preamble schemes over fading channels. Furthermore, the proposed preamble can also be used for channel estimation and obtain accurate timing synchronization with good performance. Theoretical analyses and simulations indicate that the proposed preamble scheme has superior performance to the conventional preamble schemes in various seniors. Although the proposed preamble is designed for broadcasting, its direct application in other communication scenarios is expected.

\section*{Acknowledgement}
The authors would like to thank Dr. Jean-Yves Chouinard and anonymous reviewers for the valuable comments, which strengthen the manuscript significantly. Meanwhile, authors would like to thank Dr. Jian Song and Dr. Linglong Dai for the constructive suggestions when revising this paper.

\end{document}